# Technical Preprint: Rationale and Design of a Planned Observational Study to Evaluate the Impact of Hydrocodone Rescheduling on Opioid Prescribing After Surgery

Mark D. Neuman, MD, MSc (1,2,3,4); Sean Hennessy, PharmD, PhD (2,4, 5, 6); Dylan Small, PhD (2, 4,6,7); Colleen Brensinger, MS (6); Craig Newcomb, MS (6); Lakisha Gaskins, MHS (1, 3); Duminda Wijeysundera, MD, PhD (8); Brian T. Bateman, MD, MSc (9,10); Hannah Wunsch, MD, MSc (11,12)

(1) Department of Anesthesiology and Critical Care, Perelman School of Medicine, University of Pennsylvania Perelman School of Medicine; (2) Leonard Davis Institute of Health Economics, University of Pennsylvania; (3) Center for Perioperative Outcomes Research and Transformation, University of Pennsylvania Perelman School of Medicine; (4) Center for Pharmacoepidemiology Research and Training, University of Pennsylvania Perelman School of Medicine; (5) Department of Biostatistics, Epidemiology, and Informatics, University of Pennsylvania Perelman School of Medicine; (6) Center for Clinical Epidemiology and Biostatistics, University of Pennsylvania Perelman School of Medicine; (7) Department of Statistics, The Wharton School, University of Pennsylvania; (8) Department of Anesthesia, St. Michael's Hospital and University of Toronto; (9) Department of Anesthesia, Perioperative, and Pain Medicine, Brigham and Women's Hospital and Harvard Medical School; (10) Division of Pharmacoepidemiology and Pharmacoeconomics, Department of Medicine, Brigham and Women's Hospital and Harvard Medical School; (11) Department of Critical Care Medicine, Sunnybrook Health Sciences Centre, Toronto, Ontario, Canada; (12) Department of Anesthesia and Interdepartmental Division of Critical Care Medicine, University of Toronto, Toronto, Ontario, Canada.

**Corresponding Author Address:**
Mark D. Neuman, MD, MSc
University of Pennsylvania Department of Anesthesiology and Critical Care
308 Blockley Hall
423 Guardian Drive
Philadelphia PA, 19106
Mark.neuman@uphs.upenn.edu
Tel: 215 746-7468

**ABSTRACT**

In October 2014, the US Drug Enforcement Agency (DEA) reclassified hydrocodone from Schedule III to Schedule II of the Controlled Substances Act, resulting in a prohibition on refills in the initial prescription. While this schedule change was associated with overall decreases in the rate of filled hydrocodone prescriptions and opioid dispensing, available studies conflict regarding its impact on acute opioid prescribing among surgical patients. Here, we present the rationale and design of a planned study to measure the effect of hydrocodone rescheduling using a difference-in-differences design that leverages anticipated variation in the relative impact of this policy on patients treated by surgeons that more or less frequently prescribed hydrocodone products versus other opioids prior to the schedule change. Additionally, we present findings from preliminary study conducted on a subset of our full planned sample to assess for potential differences in outcome trends over the 3 years prior to rescheduling among patients treated by surgeons who commonly prescribed hydrocodone versus those treated by surgeons who rarely prescribed hydrocodone.

**INTRODUCTION**

In October 2014, the US Drug Enforcement Agency (DEA) reclassified hydrocodone, a commonly used opioid analgesic, from Schedule III to Schedule II of the Controlled Substances Act.[1] As a result of this change, hydrocodone prescriptions written after October 2014 were newly subject to regulations that aligned with existing prescribing rules for most other opioids. Most notably, this included a prohibition on prescribing refills in the initial prescription. At a population level, the hydrocodone schedule change was associated with marked decreases in the rate of filled hydrocodone prescriptions[2, 3] as well as decreases in overall rates of opioid dispensing.[4]

Despite this, the impact of the 2014 rescheduling of hydrocodone on patterns of acute opioid prescribing among surgical patients remains unclear. One report observed initial postoperative opioid prescriptions to contain greater quantities of opioids after versus before rescheduling, potentially indicating a negative unintended consequence of hydrocodone rescheduling by which restrictions on opioid refills encouraged larger initial prescriptions.[5] A subsequent analysis failed to replicate these findings.[6] Moreover, as past analyses of postoperative opioid prescribing after versus before hydrocodone rescheduling have not controlled for secular prescribing trends, they may be limited in their ability to identify causal relationships between this policy change and prescribing outcomes. Finally, past work has not examined the impact of hydrocodone rescheduling on important outcomes that may be influenced by changes in acute opioid prescribing patterns, such as the rate of long-term opioid use following surgery.

Understanding the impact of hydrocodone rescheduling on short- and long-term opioid prescribing after surgery has major relevance to US health policy. Preventing excess opioid prescribing for acute indications, such as pain treatment after surgery, is a major target for policy interventions,[7] and mitigating the development of new long-term opioid use represents an important aspect of the rationale underlying such efforts. To the extent that the 2014 rescheduling of hydrocodone may have altered surgeons' prescribing behaviors in the acute setting through impacts on the initial surgical prescription or by limiting access to additional opioids via refills, it provides a unique opportunity to examine potential effects of changes in acute opioid prescribing behaviors on long-term opioid use outcomes.

Given these considerations, we propose to test the association of the DEA's 2014 rescheduling of hydrocodone with persistent postoperative opioid use within a sample of commercially insured US adults undergoing common general or orthopedic surgical procedures. To account for secular trends in prescribing around the time of the schedule change, we will use a difference-in-differences approach that will compare prescribing outcomes before versus after the schedule change across groups of patients who are more or less likely to experience the direct effects of this change based on their surgeon's prior tendency to prescribe hydrocodone compared to other opioids for postoperative pain management. We hypothesize that, among patients treated by surgeons who commonly prescribed hydrocodone prior to the schedule change compared to those treated by surgeons who rarely or never prescribed hydrocodone, rescheduling led to a decrease in the rate of new persistent opioid use between 90 and 180 days after surgery due to decreases in opioid exposure in the immediate postoperative period.

This technical preprint describes the data, research design, and analytic plan for our planned study. Additionally, it presents results from a preliminary analysis that used data on a subset of patients to be included in the final planned study sample who underwent surgical procedures prior to publication of the rescheduling final rule in August 2014. The primary purpose of this preliminary analysis was to assess differences in outcome trends prior to hydrocodone rescheduling among groups of patients treated by surgeons who commonly versus rarely prescribed hydrocodone as a means of assessing for evidence of potential violations of a key assumption of our planned difference-in-differences analysis, namely the presence of parallel pre-implementation outcome trends.[8, 9]

**METHODS**

**Overview of planned study design**

To estimate the effect of hydrocodone rescheduling on opioid prescribing and patient outcomes, we plan to undertake a difference-in-difference analysis that will divide patients into "exposed" and "unexposed" groups based on the anticipated impact that rescheduling would have on their care. To carry this out, we plan to take advantage of two key attributes of clinician opioid prescribing for surgery and other acute conditions in the US, namely: (1) that selection and dosage of opioids prescribed for pain treatment after a given surgery varies markedly across providers,[10, 11] and (2) prescribing habits of individual providers tend to be stable over time.[12-14] Therefore, we posit that clinicians who rarely prescribed hydrocodone products after surgery would have been minimally impacted by hydrocodone rescheduling; in contrast, we hypothesize that rescheduling would have had a marked impact on practice among those clinicians who commonly prescribed hydrocodone after surgery. Our analysis will thus estimate the impact of hydrocodone rescheduling by comparing opioid prescribing patterns before and after hydrocodone rescheduling among patients cared for by clinicians who frequently prescribed hydrocodone prior to the schedule change ("exposed") versus those treated by clinicians who rarely or never prescribed hydrocodone ("unexposed.") We reason that anticipatory changes in prescribing behavior may have taken place between the publication of the final rule regarding hydrocodone rescheduling on August 22, 2014 and the effective date of the rule on October 6, 2014;[1] therefore, we will consider dates prior to the publication of the final rule to represent the "pre-implementation" period, and dates on or after the rule's effective date to represent the "post-implementation" period (**Figure 1**).

**Data source**

We will use data from the Optum® de-identifed Clinformatics® Data Mart Database, a US health insurance database with >15 million enrollees annually. The database is geographically diverse representing all US states. Compared to the insured US population overall, Optum has a lower proportion of older adults and is less racially and ethnically diverse, with fewer individuals of black race and

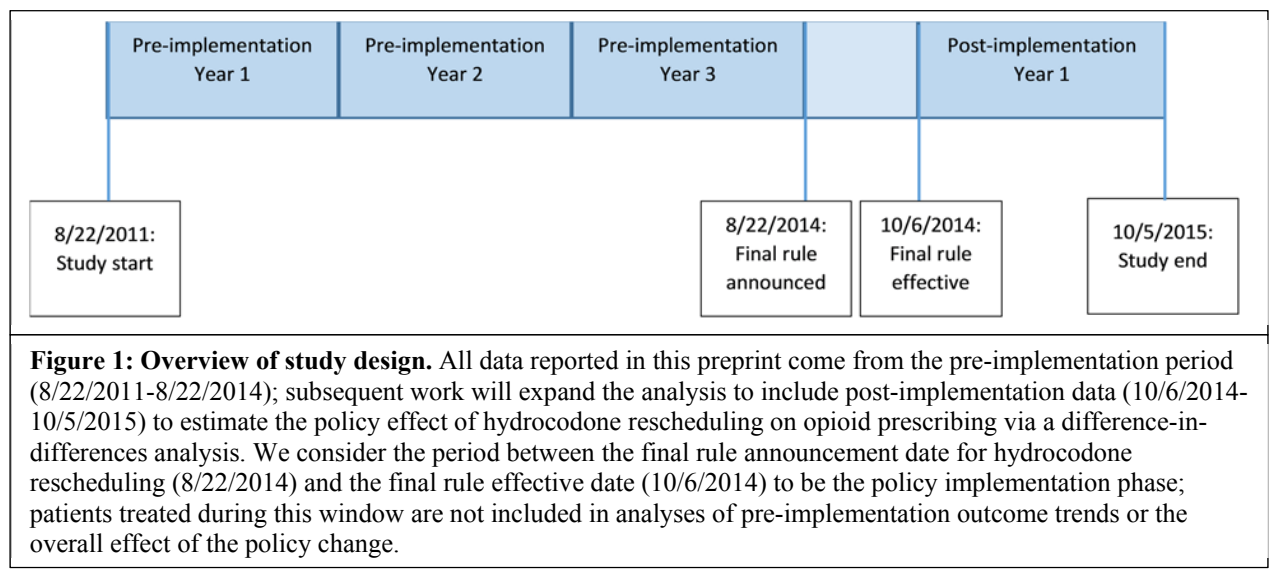

**Figure 1: Overview of study design.** All data reported in this preprint come from the pre-implementation period (8/22/2011-8/22/2014); subsequent work will expand the analysis to include post-implementation data (10/6/2014-10/5/2015) to estimate the policy effect of hydrocodone rescheduling on opioid prescribing via a difference-in-differences analysis. We consider the period between the final rule announcement date for hydrocodone rescheduling (8/22/2014) and the final rule effective date (10/6/2014) to be the policy implementation phase; patients treated during this window are not included in analyses of pre-implementation outcome trends or the overall effect of the policy change.

Hispanic ethnicity, but has comparable annual salary and education level. This study has been exempted from review by the institutional review board of the University of Pennsylvania.

**Characterizing physician prescribing patterns prior to hydrocodone rescheduling**

To characterize groups of clinicians that differ in terms of hydrocodone prescribing after surgery, we will identify all individual providers or group practices that submitted 5 or more claims between August 22, 2011 through August 21, 2014 for provision of at least one of 10 common orthopedic or general surgical procedures with an associated opioid prescription filled by the patient within 7 days of the procedure. Eligible procedures were identified based on Current Procedural Terminology (CPT) codes in physician claims and included: laparoscopic cholecystectomy; open cholecystectomy; inguinal hernia repair; laparoscopic appendectomy; open appendectomy; breast excision; carpal tunnel release; knee arthroscopy with or without meniscectomy; total knee replacement; and total hip replacement (**Table 1**). Relevant opioids included oral analgesic formulations of codeine, hydrocodone, hydromorphone, levorphanol, meperidine, morphine, oxycodone, oxymorphone, pentazocine, tramadol, fentanyl, and tapentadol. For each identified individual provider or group practice, we next calculated the proportion of all initial opioid prescriptions written within 7 days of any of our 10 study procedures that was for a hydrocodone-containing product. For the purposes of our analysis, we will categorize providers for whom hydrocodone products constituted 75% or more of initial opioid prescriptions as "hydrocodone prescribers;" we categorized providers for whom hydrocodone products constituted 25% or fewer of such prescriptions as "hydrocodone non-prescribers."

| **Table 1:** Procedures included in the study sample | |
|---|---|
| 1. Carpal Tunnel Release | 64721; 29848 |
| 2. Laparoscopic Cholecystectomy | 47562; 47563; 47564 |
| 3. Open Cholecystectomy | 47600; 47605; 47610 |
| 4. Inguinal Hernia Repair | 49505; 49507; 49520; 49521; 49525 |
| 5. Knee Arthroscopy—Meniscectomy & other | 29881; 29880; 29877; 29875; 29876; 29870 |
| 6. Total Knee Replacement | 27446;27447; 27486; 27487 |
| 7. Total Hip Replacement | 27130; 27132* |
| 8. Laparoscopic Appendectomy | 44970 |
| 9. Open appendectomy | 44950; 44960 |
| 10. Breast excision | 19301, 19302, 19120 |
| *Excluding any patient with an International Classification of Diseases 9th Revision, Clinical Modification diagnosis code indicating hip fracture (820.00-820.9) | |

**Defining the study sample**

The full analytic sample will include all patients 18 years of age and older with a submitted claim in the 3 years prior to the publication of the final rule regarding hydrocodone rescheduling (i.e. August 22, 2011 through August 21, 2014) or the year after the effective date of the schedule change (October 6, 2014 through October 5, 2015), based on the date of the procedure or hospital discharge, whichever came later. For patients with more than one eligible surgery during the window, we will use the first procedure; patients with claims for more than one eligible procedure on the same day were excluded from the sample. Next, we will exclude patients for whom the performing provider falls outside the two categories described above (i.e. hydrocodone prescribed in 25% or fewer or 75% or more of all postoperative prescriptions), such that all patients in the sample received treatment from either a hydrocodone prescriber or a hydrocodone non-prescriber. To permit a uniform window for assessment of pre-surgery comorbidities and postoperative prescribing outcomes, we will restrict the sample to patients who had (a) at least 90 days of continuous enrollment prior to the procedure or index admission date (whichever came first) and (b) at least 180 days of enrollment after the procedure or index discharge date (whichever came last). Since the hypothesized effect of hydrocodone rescheduling would be related to new restrictions on refills following initial prescriptions, we will further restrict the sample to patients who filled a prescription for an opioid within 7 days following the surgical procedure. Finally, to ensure that our outcome measures captured new long-term opioid use as opposed to the continuation of established chronic use, we will restrict our sample to those patients with no filled opioid prescriptions in the 90 days prior to surgery (i.e. "opioid-naïve" individuals).

**Outcome**

Our primary outcome will be the incidence of any opioid prescription between 90 and 180 days after surgery among previously opioid-naïve individuals.[16,17, 18] To assess mechanisms underlying any observed changes in the 30-day MME findings above, we will examine as secondary outcomes (1) the total amount of opioid (in MME) in the first postoperative prescription within 7 days of surgery or discharge, as measured in morphine milligram equivalents (MME), which we calculate using standard tables;[15] (2) the incidence of any opioid refill in the first 30 days after surgery; and (3) the total amount of opioid across all prescriptions within the first 30 days after surgery or discharge.

**Covariates**

Demographics were collected from insurance registration files. We defined baseline comorbidities using all ICD-9-CM inpatient and outpatient codes in the 180 days prior to surgery using standard crosswalks based on algorithms published by Elixhauser and colleagues.[19, 20] We created indicator variables corresponding to each of our 10 focal study procedures, as well as a variable to indicate whether the relevant provider type was an individual practitioner or a group practice. Finally, we assessed whether the index surgical procedure was performed on an ambulatory vs inpatient basis, and created a variable to characterize the length of hospital stay among individuals undergoing inpatient procedures.

**Planned Statistical Analysis**

For each patient in the sample, we defined (a) a "pre-post" variable that indicated whether their date of surgery or hospital discharge fell before the date of final rule publication (August 22, 2014) vs on or after the rescheduling effective date (October 6, 2014) and (b) an "exposed-unexposed" variable indicating whether their treating physician was classified as a hydrocodone prescriber vs a hydrocodone non-prescriber based on relevant cases performed in the 3 years prior to hydrocodone rescheduling.

Initial analyses will compare patients treated by hydrocodone prescribers vs hydrocodone non-prescribers ("exposed" vs "unexposed") with respect to patient characteristics, type of surgical procedure, and prior opioid use using standardized mean differences for continuous variables and standardized differences in proportion for categorical variables. We will graphically explore changes in outcomes relative to the date of hydrocodone rescheduling by plotting each outcome by group in 3 month intervals in the 3 years prior to the schedule change and the year following the schedule change.

We will fit regression models to predict each of the above-specified outcomes. For binary outcomes, we will use logistic regression; as we expected the distribution of MME values to be skewed, analyses of MME-based outcome variables will use generalized linear models with a gamma distribution. For a given outcome, each model predicts that outcome as a function of an interaction between the "pre-post" indicator and the "exposed-unexposed" indicator while controlling for all of the above covariates. We will consider an interaction term in this model that was significantly different from zero to indicate an effect of hydrocodone rescheduling on the relevant outcome, using $P<0.05$ as our threshold for determining statistical significance. The magnitude of the policy effect on each outcome will be assessed based on the regression coefficient for the interaction between "pre-post" and "exposed-unexposed." All models will use robust standard errors to account for clustering of observations within providers.[21] Analyses are conducted with SAS software version 9.4 (SAS Institute, Cary, NC).

**Preliminary analyses:  testing for parallel pre-implementation trends**

A key assumption of difference-in-difference analyses is that trends in outcomes for exposed and unexposed groups are the same prior to the intervention or policy change under evaluation.[8, 9] As a prerequisite to our main study, we undertook a preliminary analysis within a subset of our full planned study sample to assess for potential violations of this "parallel trends assumption" by examining the trends for all outcomes among patients treated by hydrocodone prescribers versus hydrocodone non-prescribers over the 3 years prior to the publication of the final rule for hydrocodone rescheduling in August 2014. Of note, this period corresponds to the "pre-implementation" phase of the planned analysis described above. Subsequent work will examine the data presented here plus data on patients treated in the 12 months following the effective date of hydrocodone rescheduling (i.e. the "post-implementation" period) to estimate the policy impact of hydrocodone rescheduling on opioid prescribing patterns after surgery.

Our preliminary analysis used inclusion and exclusion criteria and definitions for exposures, outcomes, and covariates as described above. Initial analyses used descriptive statistics to characterize patients treated by hydrocodone prescribers versus non-prescribers over the pre-implementation period. We examined trends in each of our outcomes among the exposed and unexposed prior to policy implementation via two methods. First we plotted and visually inspected trends for each of our 4 study outcomes over the 3 years prior to the hydrocodone rescheduling final rule publication date, by exposure group. Next, we regressed each outcome on the interaction of the binary exposure status variable (i.e. treatment by a hydrocodone prescriber vs. treatment by a hydrocodone non-prescriber) with indicators for each of the 3 years preceding the policy change among procedures performed prior to the final rule publication date, controlling for all above covariates. Analyses of binary outcomes used logistic regression; analyses of MME-based outcome variables used generalized linear models with a gamma distribution. We considered a finding of statistical significance for the interaction of exposure group with time in the period prior to the exposure to suggest a potential violation of the parallel trends assumption based on differences in trends in outcomes across the exposure and control group. As is planned for our main analysis, all models used robust standard errors to account for clustering of observations within providers and were conducted SAS software version 9.4 (SAS Institute, Cary, NC).

**RESULTS**

*Characterizing clinician prescribing patterns prior to hydrocodone rescheduling*

From August 22, 2011 through August 22, 2014, we identified 443,676 patients who received relevant procedures performed by 104,868 individual providers or group practices. Of these, 12,366 providers had at least 5 eligible procedures over the period with an associated opioid prescription within 7 days of the procedure date. Across all providers, the median proportion of hydrocodone out of postoperative opioid prescriptions was 60% (interquartile range 29%, 83%). There were 4,620 providers, including 3,659 individual providers and 961 group practices who prescribed hydrocodone for at least 75% of all postoperative prescriptions and were classified as hydrocodone prescribers. 2,798 providers, including 2,275 individual providers and 523 group practices prescribed hydrocodone for 25% or fewer of all postoperative prescriptions and were classified as hydrocodone non-prescribers.

*Defining the exposed and unexposed cohorts*

Following exclusions, we identified 52,127 patients who underwent a relevant procedure by a hydrocodone prescriber or a hydrocodone non-prescriber within the 3 years prior to the hydrocodone schedule change announcement. 33,319 (63.9%) of these cases were performed by hydrocodone prescribers and 18,808 (36.1%) were performed by non-prescribers (**Figure 2**). **Table 2** compares hydrocodone non-prescribers, those treated by hydrocodone prescribers were more often treated on an outpatient basis and more often underwent carpal tunnel release and knee arthroscopy procedures. Total joint replacement procedures were more common among patients treated by hydrocodone non-prescribers. Prior to hydrocodone rescheduling, the initial opioid prescription filled within 7 days after surgery was for a hydrocodone product in 89.8% of those treated by hydrocodone prescribers, versus 11.5% of those treated by hydrocodone non-prescribers.

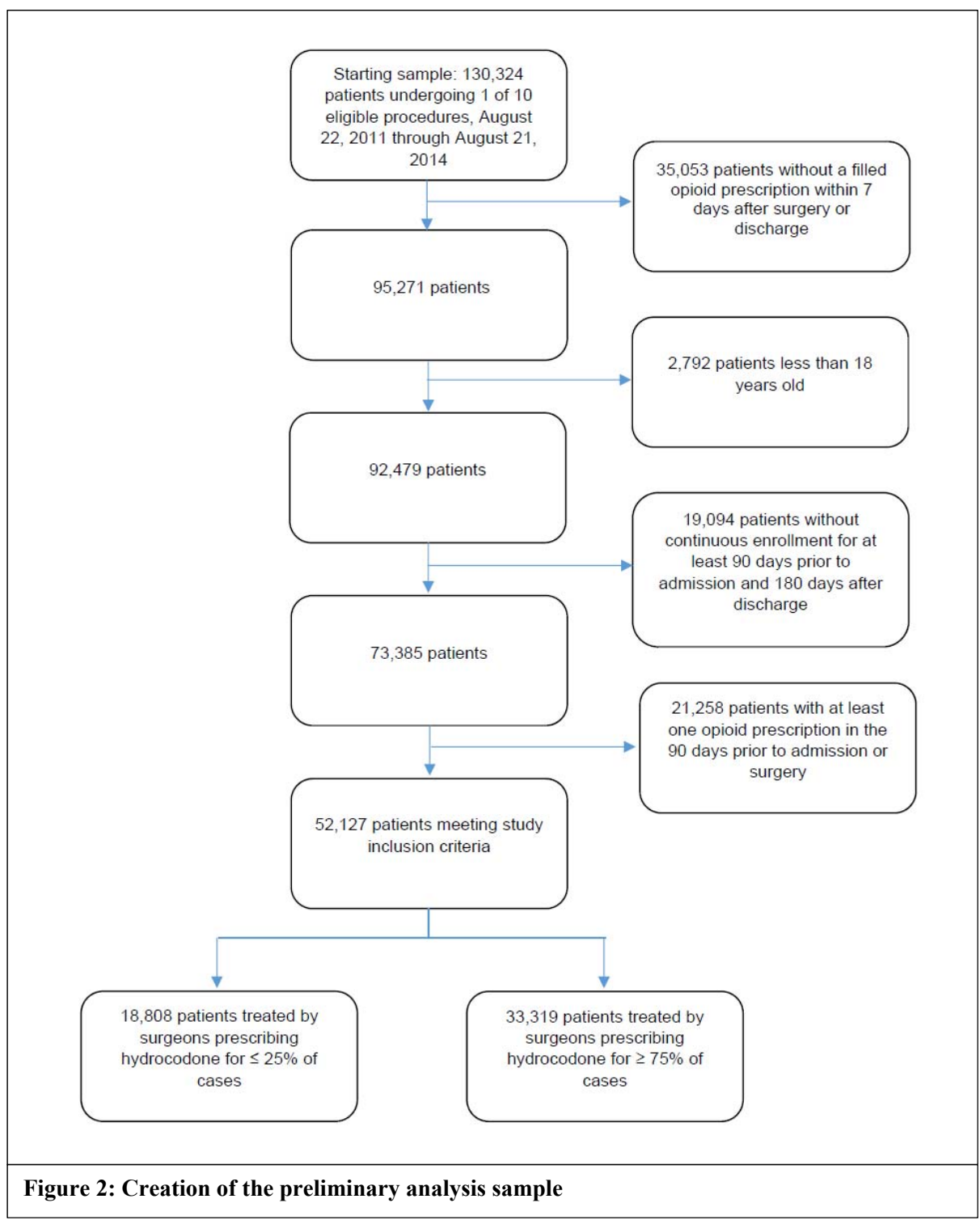

**Figure 2: Creation of the preliminary analysis sample**

| **Table 2. Characteristics of patients included in analyses of outcome trends prior to announcement of hydrocodone rescheduling in August 2014.** | | |
|---|---|---|
| Covariate | **Patients treated by surgeons prescribing hydrocodone in ≤ 25% of eligible cases (N=18,808)** | **Patients treated by surgeons prescribing hydrocodone in ≥ 75% of cases (N=33,319)** |
| Age, Median (IQR) | 54.0 (42.0-64.0) | 52.0 (40.0-62.0) |
| **Sex, N (%)** | | |
| Male | 8,700 (46.3) | 15,221 (45.7) |
| Female | 10,102 (53.7) | 18,094 (54.3) |
| **Provider type, N (%)** | | |
| Individual | 12,894 (68.6) | 21,508 (64.6) |
| Group practice | 5,914 (31.4) | 11,811 (35.4) |
| **Length of stay, N (%)** | | |
| 0 days | 12,502 (66.5) | 26,709 (80.2) |
| 1or 2 days | 3,131 (16.6) | 3,112 (9.3) |
| 3 or more days | 3,175 (16.9) | 3,498 (10.5) |
| **Procedure type, N (%)** | | |
| Laparoscopic cholecystectomy | 4,308 (22.9) | 8,433 (25.3) |
| Open cholecystectomy | 106 (0.6) | 189 (0.6) |
| Laparoscopic appendectomy | 1,646 (8.8) | 2,866 (8.6) |
| Open appendectomy | 175 (0.9) | 234 (0.7) |
| Inguinal hernia repair | 2,360 (12.5) | 3,772 (11.3) |
| Carpal tunnel release | 1,049 (5.6) | 3,814 (11.4) |
| Knee arthroscopy | 3,114 (16.6) | 7,420 (22.3) |
| Total knee replacement | 2,937 (15.6) | 2,518 (7.6) |
| Total hip replacement | 1,615 (8.6) | 1,094 (3.3) |
| Breast excision | 1,498 (8.0) | 2,979 (8.9) |
| **Comorbidities, N(%)** | | |

| Table 2. Characteristics of patients included in analyses of outcome trends prior to announcement of hydrocodone rescheduling in August 2014. | | |
|---|---|---|
| **Covariate** | **Patients treated by surgeons prescribing hydrocodone in ≤ 25% of eligible cases (N=18,808)** | **Patients treated by surgeons prescribing hydrocodone in ≥ 75% of cases (N=33,319)** |
| Congestive heart failure | 400 (2.1) | 592 (1.8) |
| Cardiac arrhythmia | 1,706 (9.1) | 2,810 (8.4) |
| Cardiac valve disease | 862 (4.6) | 1,165 (3.5) |
| Peripheral vascular disorders | 566 (3.0) | 927 (2.8) |
| Hypertension, uncomplicated | 7,416 (39.4) | 12,227 (36.7) |
| Hypertension, complicated | 601 (3.2) | 883 (2.7) |
| Other neurological disorders | 306 (1.6) | 498 (1.5) |
| Chronic pulmonary disease | 2,338 (12.4) | 3,587 (10.8) |
| Diabetes, uncomplicated | 2,289 (12.2) | 3,983 (12.0) |
| Diabetes, complicated | 455 (2.4) | 844 (2.5) |
| Hypothyroidism | 2,367 (12.6) | 3,932 (11.8) |
| Renal Failure | 499 (2.7) | 790 (2.4) |
| Liver disease | 1,239 (6.6) | 2,316 (7.0) |
| Solid tumor without metastasis | 1,545 (8.2) | 2,642 (7.9) |
| Rheumatoid arthritis | 584 (3.1) | 879 (2.6) |
| Coagulopathy | 320 (1.7) | 411 (1.2) |
| Obesity | 2,781 (14.8) | 4,191 (12.6) |
| Fluid and electrolyte disorders | 1,077 (5.7) | 1,566 (4.7) |
| Deficiency anemia | 518 (2.8) | 696 (2.1) |
| Depression | 2,089 (11.1) | 3,309 (9.9) |
| Antidepressant receipt in last 90 days | 2,716 (14.4) | 5,128 (15.4) |

*Measuring the effect of hydrocodone rescheduling on the study outcomes*

**Figure 3** depicts trends over time for each of the study outcomes prior to hydrocodone rescheduling; visual inspection of pre-implementation trends by group did not reveal marked differences in outcome rates over time. In adjusted models examining data from the pre-rescheduling period only, the interaction of the binary exposure status variable with the year variable was non-significant for receipt of any opioid between 90 and 180 days after surgery (P=0.22); total MME received in the first 30 days after surgery (P=0.29); and receipt of any refill within 30 days (P=0.75). The interaction of the exposure status variable with the year indicator was significant for total MME received within 7 days of surgery (P=0.005). The coefficient for the interaction of exposure status with the 3rd (most recent) pre-announcement year versus the 1st (earliest) year was -10.9 MME (95% CI -19.6, -2.2, P=0.014). The coefficient for the interaction of exposure status with the 2nd (intermediate) pre-announcement year (vs year 1) was -13.9 MME (95% CI -22.7, -5.1, P=0.002).

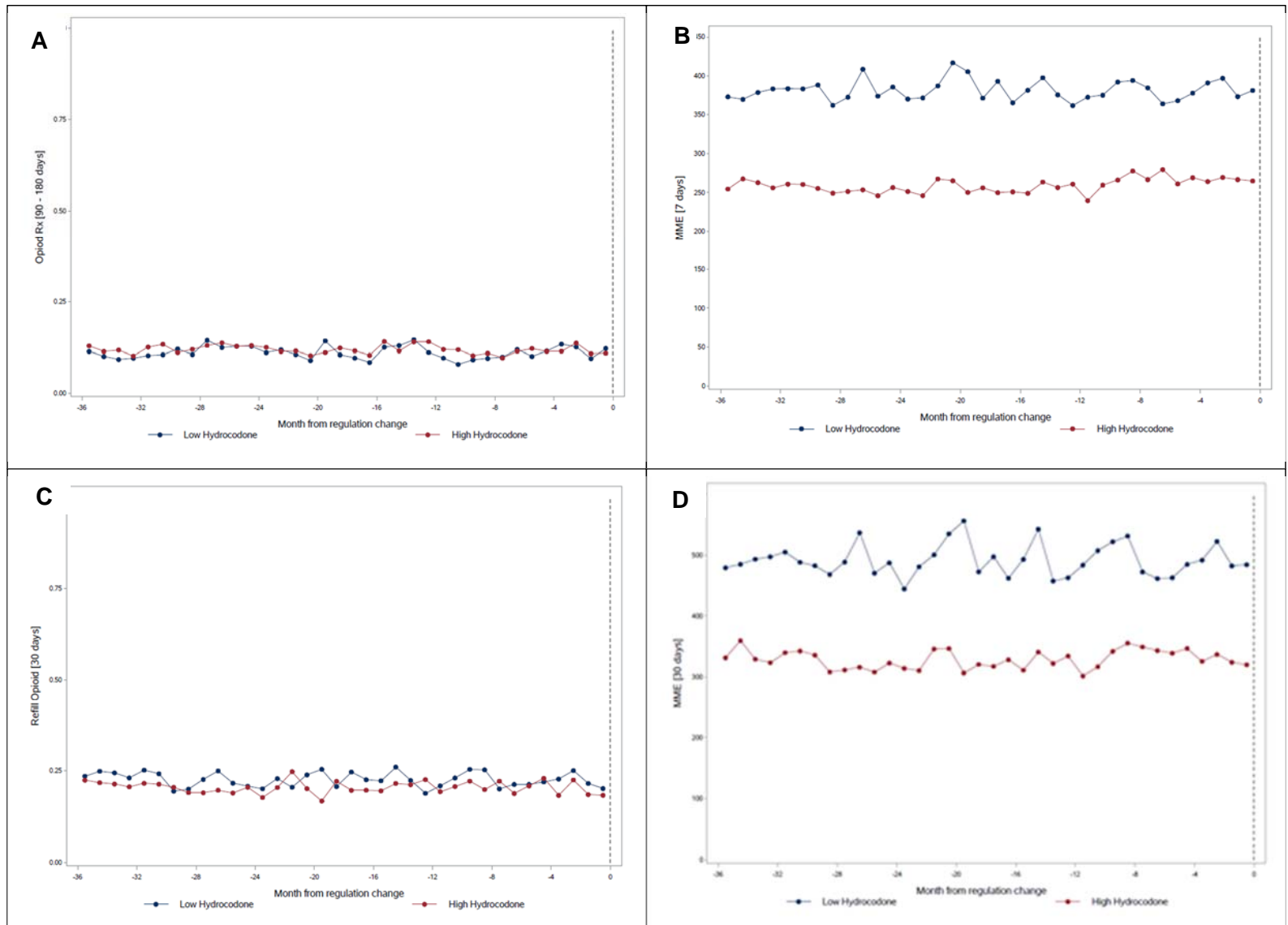

**Figure 3: Pre-implementation trends by surgeon prescriber group for (A) any filled opioid prescription between 90 and 180 days; (B) total amount of opioid dispensed in MME at 7 days; (C) any refill within 30 days of surgery; (D) total amount of opioid dispensed in Morphine Milligram Equivalents (MME) d at 30 days.** Blue lines indicate treatment by a surgeon prescribing hydrocodone in ≤ 25% of cases; red lines indicate treatment by a surgeon prescribing hydrocodone in > 75% of cases.

**DISCUSSION**

We present the rationale and design of a planned study to assess the impact of the US FDA's rescheduling of hydrocodone in 2014 from Category II to Category III of the controlled substances act on opioid prescribing practices after surgery. The proposed analysis will address key gaps in knowledge related to the impact of this policy on practice. It will employ difference-in-difference analysis, a strong inferential design, that will go beyond prior work by comparing patterns of postoperative opioid receipt before and after rescheduling among groups of patients likely to experience differential effects of hydrocodone rescheduling. Additionally, it will examine opioid prescribing outcomes beyond the immediate period after surgery to examine the downstream impact of hydrocodone rescheduling on the development of new long-term opioid use among previously opioid-naïve surgical patients. Finally, the proposed analysis will take advantage of a large and representative national claims dataset.

Analyses presented here support the feasibility and rationale of our planned analytic approach. Specifically, using data from the three years prior to the publication of the final rule on hydrocodone rescheduling, we have demonstrated that distinct comparison groups of surgical patients can be constructed based on surgeons' established prescribing patterns to preferentially prescribe or avoid hydrocodone products for postoperative pain control. Of note, we find low-risk procedures such as carpal tunnel release and knee arthroscopy constitute a greater proportion of all cases performed by hydrocodone prescribers as compared to hydrocodone non-prescribers. While it is likely that differences in prescribing patterns may relate to surgeons' individual habits or practice styles this specific finding suggests that selection of hydrocodone versus other products may also relate to anticipated pain severity. From the standpoint of our planned analysis, this observation highlights the importance of adjustment for case mix differences between exposure and control groups.

Additionally, our preliminary analysis does not find evidence for violations of the parallel trends assumption for key pre-specified outcomes for our planned analysis, namely receipt of any opioid between 90 and 180 days after surgery, total MME dispensed in the first 30 days after surgery, and receipt of any refill within 30 days. While we do observe statistically significant differences across study groups in trends over time in total MME received within 7 days after surgery, the magnitude of these differences is small in absolute terms and unlikely to be clinically significant. To assess this empirically, we will plan in the full study to conduct additional secondary analyses using estimators that account for differences in pre-exposure trends via matching or statistically controlling for such differences in the regression model.[22]

In summary, we have presented the rationale and design for a planned study to assess the impact of the US FDA's rescheduling of hydrocodone in 2014 from Category II to Category III of the controlled substances act on opioid prescribing practices after surgery. Additionally, prerequisite analyses presented here support the feasibility of the proposed approach and validate key analytic assumptions, while also highlighting important considerations for interpretation of subsequent analyses. Future work will implement this approach in the full study data set to measure the effect of the FDA hydrocodone rescheduling policy on postoperative opioid prescribing.

**ACKNOWLEDGMENTS**

This study was supported by a grant from the National Institute on Drug Abuse (#1R01DA042299-01A1)

REFERENCES

1.    US Drug Enforcement Administration. Final Rule: Schedules of Controlled Substances: Rescheduling of Hydrocodone Combination Products From Schedule III to Schedule II. 21 CFR Part 1308. Federal Register Volume 79, Number 163.  2014:49661-49682.

2.    Jones CM, Lurie PG, Throckmorton DC. Effect of US Drug Enforcement Administration's Rescheduling of Hydrocodone Combination Analgesic Products on Opioid Analgesic Prescribing. *JAMA Intern Med.* Mar 2016;176(3):399-402.

3.    Kuo YF, Raji MA, Liaw V, Baillargeon J, Goodwin JS. Opioid Prescriptions in Older Medicare Beneficiaries After the 2014 Federal Rescheduling of Hydrocodone Products. *J Am Geriatr Soc.* May 2018;66(5):945-953.

4.    Raji MA, Kuo YF, Adhikari D, Baillargeon J, Goodwin JS. Decline in opioid prescribing after federal rescheduling of hydrocodone products. *Pharmacoepidemiol Drug Saf.* May 2018;27(5):513-519.

5.    Habbouche J, Lee J, Steiger R, et al. Association of Hydrocodone Schedule Change With Opioid Prescriptions Following Surgery. *JAMA Surg.* Aug 22 2018.

6.    Tan WH, Feaman S, Milam L, et al. Postoperative opioid prescribing practices and the impact of the hydrocodone schedule change. *Surgery.* Oct 2018;164(4):879-886.

7.    Lowenstein M, Grande D, Delgado MK. Opioid Prescribing Limits for Acute Pain - Striking the Right Balance. *N Engl J Med.* Aug 9 2018;379(6):504-506.

8.    Ryan AM, Burgess JF, Jr., Dimick JB. Why We Should Not Be Indifferent to Specification Choices for Difference-in-Differences. *Health Serv Res.* Aug 2014;50(4):1211-1235.

9.    Dimick JB, Ryan AM. Methods for evaluating changes in health care policy: the difference-in-differences approach. *JAMA.* Dec 10 2014;312(22):2401-2402.

10. Hill MV, McMahon ML, Stucke RS, Barth RJ, Jr. Wide Variation and Excessive Dosage of Opioid Prescriptions for Common General Surgical Procedures. *Ann Surg.* Apr 2017;265(4):709-714.

11. Eid AI, DePesa C, Nordestgaard AT, et al. Variation of Opioid Prescribing Patterns among Patients undergoing Similar Surgery on the Same Acute Care Surgery Service of the Same Institution: Time for Standardization? *Surgery.* Nov 2018;164(5):926-930.

12. Brookhart MA, Wang PS, Solomon DH, Schneeweiss S. Evaluating short-term drug effects using a physician-specific prescribing preference as an instrumental variable. *Epidemiology.* May 2006;17(3):268-275.

13. Brookhart MA, Rassen JA, Wang PS, Dormuth C, Mogun H, Schneeweiss S. Evaluating the validity of an instrumental variable study of neuroleptics: can between-physician differences in prescribing patterns be used to estimate treatment effects? *Med Care.* Oct 2007;45(10 Suppl 2):S116-122.

14. Brookhart MA, Schneeweiss S. Preference-based instrumental variable methods for the estimation of treatment effects: assessing validity and interpreting results. *Int J Biostat.* 2007;3(1):Article 14.

15. US Centers for Medicare and Medicaid Services. Opioid Oral Morphine Milligram Equivalent (MME) Conversion Factors. . 2017; https://www.cms.gov/Medicare/Prescription-Drug-Coverage/PrescriptionDrugCovContra/Downloads/Opioid-Morphine-EQ-Conversion-Factors-April-2017.pdf. Accessed April 14, 2019.

16. Brummett CM, Waljee JF, Goesling J, et al. New Persistent Opioid Use After Minor and Major Surgical Procedures in US Adults. *JAMA Surg.* Jun 21 2017;152(6):e170504.

17. Johnson SP, Chung KC, Zhong L, et al. Risk of Prolonged Opioid Use Among Opioid-Naive Patients Following Common Hand Surgery Procedures. *J Hand Surg Am.* Oct 2016;41(10):947-957 e943.

18.     Lee JS, Hu HM, Edelman AL, et al. New Persistent Opioid Use Among Patients With Cancer After Curative-Intent Surgery. *J Clin Oncol.* Dec 20 2017;35(36):4042-4049.

19.     Quan H, Sundararajan V, Halfon P, et al. Coding algorithms for defining comorbidities in ICD-9-CM and ICD-10 administrative data. *Med Care.* Nov 2005;43(11):1130-1139.

20.     Elixhauser A, Steiner C, Harris DR, Coffey RM. Comorbidity measures for use with administrative data. *Med Care.* Jan 1998;36(1):8-27.

21.     White H. A Heteroskedasticity-Consistent Covariance-Matrix Estimator and a Direct Test for Heteroskedasticity. *Econometrica.* 1980;48(4):817-838.

22.     Ryan AM, Kontopantelis E, Linden A, Burgess JF, Jr. Now trending: Coping with non-parallel trends in difference-in-differences analysis. *Stat Methods Med Res.* Nov 25 2018:962280218814570.